\documentclass[a4paper,brazil,12pt,nofootinbib,aps,pra]{revtex4}

\usepackage[T1]{fontenc}
\usepackage{moreverb}          

\usepackage{amsmath}  
\usepackage{slashed}

\usepackage{bm}       
\usepackage{braket}   
\usepackage{nicefrac} 
\usepackage{bbm}      
\usepackage{mathrsfs} 
\usepackage{multirow} 
\usepackage{cancel}

\usepackage{graphicx}
\usepackage{subfigure}

\usepackage{caption}
\DeclareCaptionFont{xipt}{\fontsize{11}{13}\mdseries}
\usepackage[font=xipt,labelfont=bf]{caption}



\usepackage{hyperref}
\hypersetup{
citebordercolor=0 .75 1,
linktocpage=true
}

\makeatletter
\renewcommand*\email[1][E-mail: ]{\begingroup\sanitize@url\@email{#1}}%
\makeatother

\begin{document}

\title{A statistical mechanical analysis on the possibility of achieving \\
fair cylindrical dice}

\author{M. N. C. Brustelo}
\email{matheus.brustelo@gmail.com}

\author{M. M. Vivaldi}
\email{mateusmvivaldi@gmail.com}

\author{F. Marques}
\email{fabriciomarques@if.usp.br}


\affiliation{University of S\~ao Paulo, Institute of Physics, 66318, 05315-970, S\~ao Paulo, SP, Brazil}

\date{\today}  
               %

\begin{abstract}{
Many have dedicated their time trying to determine the ideal conditions for a cylinder to have equal probabilities of falling with one of its faces facing upwards or on its side. However, to this day, there is no concrete analysis of what these conditions should be. In order to determine such circumstances, a theoretical analysis was conducted, considering approaches from Rigid Body Dynamics and Statistical Mechanics. An experimental system was also built to improve control over the launches, and a comparative analysis was performed between the results obtained experimentally and the theory. It was concluded that the environment and other launching conditions have a significant influence; nevertheless, it is possible, under controlled conditions, to determine, within certain limits, the expected probabilities.
  }\end{abstract}

\maketitle    

                
\newpage      




\section{\label{sec:introducao}Introduction}


When studying probability and statistics, it is common to encounter examples of a 
cubic die or of coins that can be flipped to heads or tails. 
Although one can consider cases of a biased die, where most of the mass is 
concentrated near the face opposite to the one that is more likely to end up 
facing upwards, there is no \emph{a priori} reason to 
expect that each face of the cube does not have an equal probability of $1/6$ of 
ending up facing upwards. Similarly, in the case of a coin, if a large number of 
tosses are performed, the common expectation is to obtain heads for approximately 
half the number of tosses and tails for the remaining half. 
However, in the case of the coin, we can observe an interesting fact. 
Most coins are cylinders that have a small thickness 
$H$ compared to their radius $R$. It is possible, with some care and patience, to 
balance a coin with its side resting on a horizontal flat surface, so that it is 
neither in the ``heads'' nor ``tails'' state, but rather in a third state, which we 
can call the ``side'' state. Much less likely, obviously, is for the coin to end up 
in this position after an arbitrary toss.

Just as the idea of a coin landing in the ``side'' state after a toss seems 
extremely improbable, it may also appear nonsensical to expect that a stick, which 
is nothing more than a cylinder with $H \gg R$, ends up in a state equivalent to 
``heads'' or ``tails'', i.e., with one of its faces facing downwards, after being 
thrown. Unlike the case of the coin, therefore, the common expectation is for the 
stick to end up in the ``side'' state when landing on a flat and horizontal surface.

The combination of the common expectations for the coin and the stick, therefore, 
leads us to the question proposed in the statement of Problem 7 of the 
35th International Young Physicists' Tournament 2022 
(IYPT 2022)~\cite{IYPT,IYPT2022} :

\begin{quote}
  \emph{  ``To land a coin on its side is often associated with the idea of a rare 
  occurrence. What should be the physical and geometrical characteristics of a 
cylindrical dice so that it has the same probability to land on its side and one of 
its faces?''}
\end{quote}

In other words, for a cylinder, the probability of obtaining ``face'' is $P_F 
\approx 1$, and the probability of obtaining ``side'' is $P_S \approx 0$ for $H \ll 
R$ (coin). On the other hand, these same probabilities become $P_F \approx 0$ and 
$P_S \approx 1$ for $H \gg R$ (stick). However, there could be some ideal $H/R$ 
ratio that, combined with a certain set of specific physical conditions, would make 
$P_S = 1/3$ and $P_F = 2 P_{F_1} = 2 P_{F_2} = 2/3$. If the right set of conditions 
could be determined, we would then have a ``fair cylindrical die''.

Besides the specific problem proposed in IYPT 2022, which requires the analysis of a 
cylindrical die, the idea could, in principle, be generalized to analyze the 
conditions that allow the construction of fair dice with different geometries. In 
role-playing games, for example, it is common to use dice with shapes different from 
the typical cubic form. An interesting mathematical analysis
concerning the alowed forms for convex polyedra to the considered suitable
candidates to constitute ``fair dice'' can be found in
reference~\cite{Diaconis}. That study, though, does not take into account the
physics of a launch, but instead will focus on symmetry group arguments. A
comment is made about other fair polyedra, including a solid produced by cutting
off the tips of a di-pyramid with $2n$ identical triangular faces with two
planes parallel to its base and equidistant from it. Following that comment,
they point out that the location of those cuts would possibly depend on
mechanical properties of such a die and also of the surface where it will land.
To justify that statement, they cite an early paper~\cite{Keller} that 
proposes to analyze the motion of a tossed coin in order to seek for its
conections with the probabilities of getting heads or tails, but restricted to
the situation of vertical lauches with landings on a plane surface which 
completely absorbs the impact, as sand or mud.
In this context, our cylinders can be condirered an extended version of the
cutted di-pyramids described by \cite{Diaconis} in a limit in which $n$ and the
height of the di-pyramid are taken to be infinite, but the two cuts are made at 
a finite distance from each other, which will be the height $H$ of the cylinder. 
Also, instead of restricting ourselves to a complex yet restrictive study on 
the mechanics of a cylinder's motion, we will move 
towards a statistical mechanical analysis, therefore presenting a different
perspective about this matter. 

This work is organized as follows. Section~\ref{sec:fundamentos} will present some 
of the technical aspects of the statistical interpretation of results from 
cylindrical dice rolls, as well as the theory associated with the motion of a 
cylindrical rigid body. In Section~\ref{sec:descricao}, a general description of the 
experimental apparatus used will be provided. Next, in Section~\ref{sec:analise}, 
the obtained data will be presented along with their respective analysis and 
discussion. Finally, in Section~\ref{sec:conclusao}, the concluding remarks will be 
presented.



\section{\label{sec:fundamentos}Theoretical Foundations}

\subsection{Free motion of a cylinder falling under the action of gravity}

The position $\vec{R}$ of the Center of Mass (CM) of a rigid body with respect to an 
inertial reference frame is affected by external forces acting on the body. In the 
case of the cylindrical die, during its flight motion, the weight force and air 
resistance forces come into play. For sufficiently small cylinders, launched at low 
heights to avoid reaching high velocities, air resistance forces can be neglected as 
a first approximation. These forces, in addition to affecting the CM position, could 
also influence the angular velocity of the cylinder, particularly around an axis 
contained in the plane defined by the principal axes $x_1$ and $x_2$. 

\begin{figure}[h!]
    \begin{center}
        \includegraphics[width=0.70\textwidth]{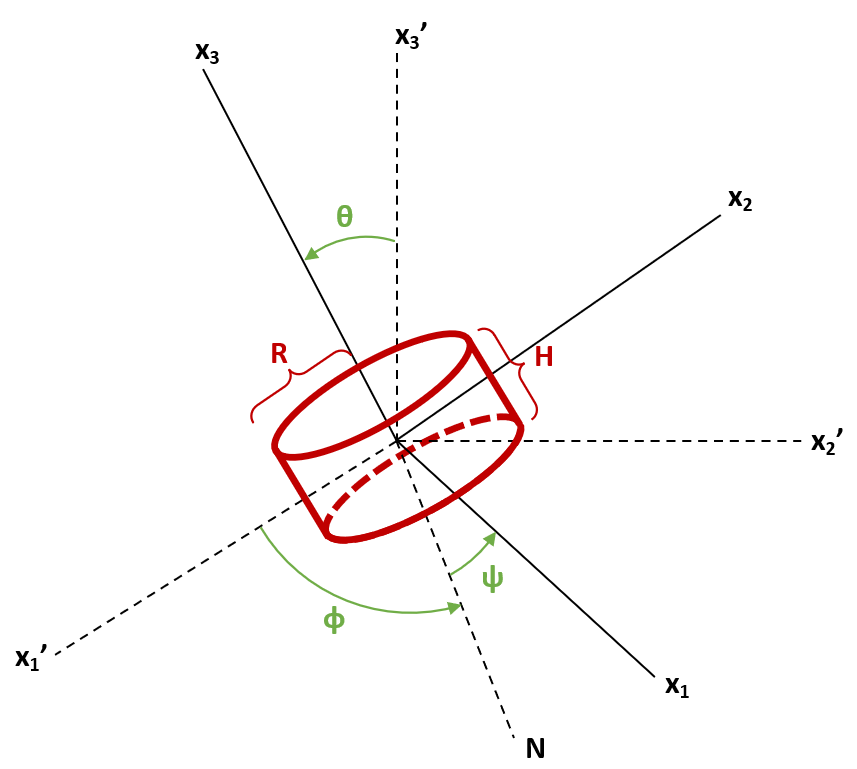}
    \end{center}
    \caption{Fixed orientation system $S^{\prime}$, body's principal system $S$, and the Euler angle convention to be used.
    \label{ConvencaoDeAngulosDeEuler}
    }
\end{figure}

In this regard, a theory that does not consider the air resistance forces is limited 
in not taking into account the aerodynamics of the cylinder's motion, which can make 
the initial collision of ``face'' or ``side'' more or less probable. However, unless 
a set of materials is chosen for the cylinder and surface such that the cylinder 
sticks to the surface upon collision, the final state will only occur after a 
sequence of multiple collisions. On the other hand, the weight force acts as if it 
were concentrated entirely on the CM, not affecting the rotational motion of the 
cylinder during its flight.

Therefore, in the approximation that neglects the dissipative forces exerted by the 
contact with air, we can analyze the flight motion of the cylindrical die as the 
free rotational motion of a rigid body whose CM translates by inertia in the 
horizontal direction and under the influence of weight in the vertical direction. In 
this motion, mechanical energy $E$ and angular momentum $l$ are conserved 
quantities. The rotational motion is governed by how mass is distributed within the 
cylinder. For a homogeneous mass distribution in a cylinder with radius $R$, height 
$H$, and mass $M$, the principal moments of inertia will be:

\begin{equation}
\label{MomentosDeInercia}
    I_1 = I_2 = I = \frac{MR^2}{4}+\frac{MH^2}{12} 
    \quad \text{and} \quad
    I_3 = \frac{MR^2}{2}
\end{equation}
and the mechanical energy can be written as:

\begin{equation}
\label{EnergiaMecanica}
    E = \frac{M V^2}{2} 
    + \frac{I_1 \omega_1^2}{2}+ \frac{I_2 \omega_2^2}{2}+ \frac{I_3 \omega_3^2}{2}
    + M g h
\end{equation}
where $V = |\dot{\vec{R}}|$ is the velocity of the CM, $h$ is the height of the CM 
relative to the level of the horizontal plane where the cylinder will land, and 
$\omega_1$, $\omega_2$, and $\omega_3$ are the components of the angular velocity 
vector of the cylinder in the principal axis system.

The terms in Equation \eqref{EnergiaMecanica} proportional to the moments of inertia 
correspond to rotational kinetic energy. Using Euler angles as defined in 
Figure~\ref{ConvencaoDeAngulosDeEuler}, the components of the angular velocity can 
be written as:

\begin{equation}
\label{ComponentesVelocidadeAngular}
    \omega_1 = \dot{\phi} \sin \theta \sin \psi + \dot{\theta} \cos \psi
    \; , \quad 
    \omega_2 =  \dot{\phi} \sin \theta \cos \psi - \dot{\theta} \sin \psi
    \quad \text{and} \quad 
    \omega_3 = \dot{\phi} \cos \theta + \dot{\psi}
\end{equation}

Substituting the expressions \eqref{ComponentesVelocidadeAngular} for the components 
of the angular velocity in \eqref{EnergiaMecanica}, considering the symmetry of the 
mass distribution that makes $I_1 = I_2 = I$, and assuming a moment immediately 
before the collision of the cylinder with the landing plane, which makes $h = R 
\left( \sin \theta + \frac{H}{2R} \cos \theta \right)$, the mechanical energy will 
be given by:

\begin{equation}
    \label{EnergiaMecanicaComAngulos}
    E = \frac{MV^2}{2}
    + \frac{I}{2} \left(\dot{\phi}^2 \sin^2 \theta + \dot{\theta}^2 \right)
    + \frac{I_3}{2} \left( \dot{\phi} \cos \theta + \dot{\psi} \right)^2
     + MgR \left( \sin \theta + \frac{H}{2R} \cos \theta \right)
\end{equation}

The angular momentum, by its turn, will have components in the principal system, 
given by:

\begin{equation*}
    l_1 = I \omega_1
    \; , \quad 
    l_2 =  I \omega_2
    \quad \text{and} \quad 
    l_3 = I_3 \omega_3
\end{equation*}
and, using the expressions \eqref{ComponentesVelocidadeAngular}, we can write the 
squared magnitude of the total angular momentum, $l^2$, which is another constant of 
the flight motion:

\begin{equation}
\label{MomentoAngularTotalQuadrado}
    l^2 = l_1^2+l_2^2+l_3^2
    = I^2 \left( \dot{\phi}^2 \sin^2 \theta + \dot{\theta}^2 \right)
    + I_3^2 \left( \dot{\phi} \cos \theta + \dot{\psi} \right)^2
\end{equation}

Changing the sign of the potential energy term in expression 
\eqref{EnergiaMecanicaComAngulos}, we obtain the Lagrangian $L$, which is cyclic in 
$\phi$ and $\psi$, i.e., it does not depend explicitly on these angles. Therefore, 
in addition to the mechanical energy $E$, the canonically conjugate momenta 
$p_{\phi}$ and $p_{\psi}$ are also constants of motion~\cite{Marion}, given by:

\begin{equation}
    \begin{aligned}
    p_{\phi} &= \frac{\partial L}{\partial \dot{\phi}} 
        = \left( I \sin^2 \theta + I_3 \cos^2 \theta \right) \dot{\phi}
        + I_3 \dot{\psi} \cos \theta
        \\
    p_{\psi} &= \frac{\partial L}{\partial \dot{\psi}}
        = I_3 \left( \dot{\phi} \cos \theta + \dot{\psi} \right) = I_3 \omega_3
    \end{aligned}
\end{equation}
and thus, we can rewrite the expression \eqref{EnergiaMecanicaComAngulos} for the 
mechanical energy as:

\begin{equation}
    \label{EnergiaMecanicaLivreFinal}
    E = 
    \frac{MV^2}{2} + \frac{I \dot{\theta}^2}{2}
    + \frac{\left( p_{\phi} - p_{\psi} \cos \theta \right)^2}{2 I \sin^2 \theta}
    + \frac{p_{\psi}^2}{2 I_3}
    + MgR \left( \sin \theta + \frac{H}{2R} \cos \theta \right)
\end{equation}

\subsection{Dynamics of the collision between a cylinder and a horizontal flat surface}

\begin{figure}[h!]
    \begin{center}
        \mbox{ 
            \subfigure[Axes and dimensions.]{
                \includegraphics[width=0.40\columnwidth]{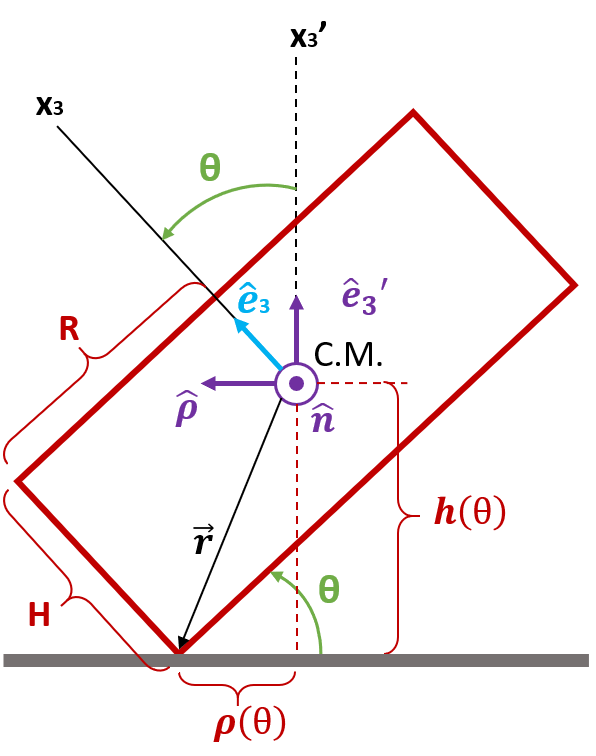}
                \label{GeometriaEixos}
                }
            \subfigure[Acting forces.]{
                \includegraphics[width=0.50\columnwidth]{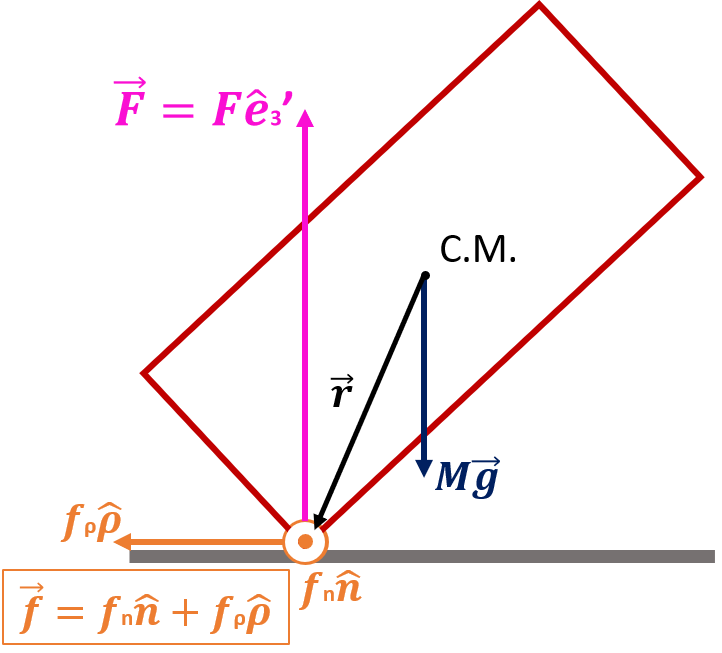}
                \label{GeometriaForcas}
                }
        }
    \end{center}
    \caption{Geometry of the cylinder system at the instant of collision with the landing plane.
    \label{GeometriaDaColisao}
    }
\end{figure}

The position of the contact point at the instant of collision can be written as:

\begin{equation}
    \label{PosicaoDoPontoDeContato}
    \vec{r} = 
    \rho(\theta) \hat{\rho}
    - h(\theta) \hat{e}_3^{\prime}
\end{equation}
where 
$\rho(\theta) = R \left( - \cos \theta + \frac{H}{2R} \sin \theta \right)$,
$h(\theta) = R \left( \sin \theta + \frac{H}{2R} \cos \theta \right)$
and 
$\hat{\rho} = \left( - \hat{e}_1^{\prime} \sin \phi + \hat{e}_2^{\prime} \cos \phi 
\right)$ is a horizontal unit vector perpendicular to the nodal line
$\hat{n} = \left( \hat{e}_1^{\prime} \cos \phi + \hat{e}_2^{\prime} \sin \phi 
\right)$. The unit vectors $\hat{\rho}$, $\hat{n}$, and $\hat{e}_3^{\prime}$ define 
an orthonormal basis in space, such that $\hat{\rho} \times \hat{n} = 
\hat{e}_3^{\prime}$,
$\hat{n} \times \hat{e}_3^{\prime} = \hat{\rho}$, and
$\hat{e}_3^{\prime} \times \hat{\rho} = \hat{n}$ (Fig.~\ref{GeometriaEixos}).

During the contact, the cylinder experiences a normal force 
$\vec{F} = F \hat{e}_3^{\prime}$ and a frictional force 
$\vec{f} = f_{\rho}\hat{\rho} + f_{n}\hat{n}$. Therefore, the torque with respect to 
the center of mass is:

\begin{equation}
    \label{TorqueNaColisao}
    \vec{\tau}^{\prime} 
    = \vec{r} \times \left( \vec{F} + \vec{f} \right)
    = -\bigg( 
        h(\theta) f_{\rho} +\rho(\theta) F 
    \bigg) \hat{n}
    + f_n \bigg( 
        h(\theta) \hat{\rho} + \rho(\theta) \hat{e}_3^{\prime} 
    \bigg)
\end{equation}

The torque from the normal force can contribute to changes in the angular frequency 
$\dot{\theta}$ and the velocity of the center of mass. The frictional torque can 
have two contributions, one tangential (in the direction of $\hat{n}$) and one 
radial (in the direction of $\hat{\rho}$) with respect to the arc that the contact 
point tends to describe around the vertical axis $x_3^{\prime}$, thus potentially 
causing changes in all three angular frequencies $\dot{\phi}$, $\dot{\theta}$, and 
$\dot{\psi}$ (Fig.~\ref{GeometriaForcas}).

The energy $W_1$ dissipated during the first collision of a given cylinder with the 
landing plane can receive a contribution $W_F$ due to the normal force $\vec{F}$, 
since the collision is not perfectly elastic, and another contribution $W_f$ due to 
the friction $\vec{f}$, provided that the cylinder slides on the plane. Therefore:

\begin{equation}
\begin{aligned}
    \label{TrabalhoDissipativo}
    W_1  &=  W_F + W_f = \int \vec{F} \cdot d\vec{r} + \int \vec{f} \cdot d\vec{r}  
    \\
    &= \int_0^{\Delta t_1} \left( \vec{F} + \vec{f} \right) \cdot \vec{v}\, dt
    = \int_0^{\Delta t_1} \left( \vec{F} + \vec{f} \right) \cdot 
    \left( \vec{V} + \vec{\omega}^{\prime} \times \vec{r} \right) dt
    \\
    &= \int_0^{\Delta t_1} \underbrace{\left( \vec{F} + \vec{f} \right)}_{\vec{F}_{\text{contact}}} \cdot \vec{V} dt 
    + \int_0^{\Delta t_1} \underbrace{\left( \vec{F} + \vec{f} \right) \cdot \left( \vec{\omega}^{\prime} \times \vec{r} \right)}_{\vec{\omega}^{\prime} \cdot \left( \vec{r} \times \vec{F}_{\text{contact}} \right) = \vec{\omega}^{\prime} \cdot \vec{\tau}^{\prime}} dt
\end{aligned}
\end{equation}
where we used the property of the triple product to simplify the expression
$ \vec{A} \cdot ( \vec{B} \times \vec{C} ) 
= \vec{B} \cdot ( \vec{C} \times \vec{A} )
= \vec{C} \cdot ( \vec{A} \times \vec{B} )$. 

The integrand of the last term in \eqref{TrabalhoDissipativo} is:

\begin{equation}
    \vec{\omega}^{\prime} \cdot \vec{\tau}^{\prime} 
    = \vec{\omega} \cdot \vec{\tau}
    = I \omega_1 \dot{\omega}_1 - \cancel{(I-I_3) \omega_1 \omega_2 \omega_3}
    + I \omega_2 \dot{\omega}_2 - \cancel{(I_3 - I)\omega_2 \omega_3 \omega_1}
    + I_3 \omega_3 \dot{\omega}_3
\end{equation}
where the covariance of scalar products under rotations was used to transition from 
the fixed orientation system ($S^{\prime}$) to the body system ($S$), and then the 
torque components were substituted using the Euler equations for the motion of rigid 
bodies~\cite{Marion}. Thus, we have that the integral in the last term is:

\begin{equation}
    \label{EnegiaResidual0}
    \begin{aligned}
   \int_0^{\Delta t_1} & \vec{F}_{\text{contact}} \cdot \left( \vec{\omega}^{\prime} \times \vec{r} \right)dt
   = \int_0^{\Delta t_1}  \vec{\omega}^{\prime} \cdot \left( \vec{r} \times \vec{F}_{\text{contact}} \right) dt
   \\ &
   = \int_0^{\Delta t_1}  \vec{\omega}^{\prime} \cdot \vec{\tau}^{\prime} dt
   = \int_0^{\Delta t_1} \left( 
   I \omega_1 \dot{\omega}_1 
    + I \omega_2 \dot{\omega}_2 
    + I_3 \omega_3 \dot{\omega}_3
   \right)
   \\ &
   = \frac{I \omega_{1,1}^{2}}{2} + \frac{I \omega_{2,1}^2}{2} + \frac{I_3 \omega_{3,1}^2}{2}
    - \frac{I \omega_{1,0}^2}{2} - \frac{I \omega_{2,0}^2}{2} - \frac{I_3 \omega_{3,0}^2}{2}
    \end{aligned}
\end{equation}

The integral in the penultimate term of \eqref{TrabalhoDissipativo} is:

\begin{equation}
\label{PenultimoTermo}
    \int_0^{\Delta t_1} \vec{F}_{\text{contact}} \cdot \vec{V} dt 
    = \int_0^{\Delta t_1} \vec{F} \cdot \vec{V} dt 
    + \int_0^{\Delta t_1} \vec{f} \cdot \vec{V} dt
\end{equation}
where $\vec{F}$ and $\vec{f}$ can be determined using the second Newton's law:

\begin{equation*}
    \frac{d \vec{p}}{dt} = \vec{F} + \vec{f} + M\vec{g} 
    \Rightarrow
    M \frac{d}{dt} \bigg\lbrace 
    V_{\rho} \hat{\rho} + V_n \hat{n} + V_3 \hat{e}_3^{\prime} 
    \bigg\rbrace 
    = f_{\rho} \hat{\rho} + f_n \hat{n} 
    + \left( F - Mg \right) \hat{e}_3^{\prime}
\end{equation*}
and noting that $\frac{d \hat{\rho}}{dt} = \omega_3^{\prime} \hat{e}_3^{\prime} 
\times \hat{\rho} = \dot{\phi} \hat{n}$ and $\frac{d \hat{n}}{dt} = 
\omega_3^{\prime} \hat{e}_3^{\prime} \times \hat{n} = - \dot{\phi} \hat{\rho}$, we 
get:

\begin{equation}
    \label{Forcas}
    F = Mg+M\frac{dV_3}{dt}
    \; , \quad
    f_{\rho} = - M\dot{\phi}V_n + M \frac{dV_{\rho}}{dt}
    \; , \quad
    f_n = M \dot{\phi}V_{\rho} + M \frac{dV_n}{dt}
\end{equation}

Thus, the integrals in \eqref{PenultimoTermo} become:

\begin{equation}
    \label{EnergiaResidual1}
    \int_0^{\Delta t_1} 
    \vec{F} \cdot \vec{V} dt
    = \int_0^{\Delta t_1} 
    \left( 
    Mg + M\frac{dV_3}{dt}
    \right) 
    V_3  dt
    = \left( \frac{MV_{3,1}^2}{2}+ Mgh_1 \right) 
    - \left( \frac{MV_{3,0}^2}{2} + Mgh_0 \right)
\end{equation}
which affects the height of the cylinder and the vertical component of the CM 
velocity after the collision, and

\begin{equation}
\label{EnergiaResidual2}
\begin{aligned}
   \int_0^{\Delta t_1} 
& 
    \vec{f} \cdot \vec{V} dt
    = \int_0^{\Delta t_1} 
    \left( 
    f_{\rho}\hat{\rho} + f_{n}\hat{n}
    \right) \cdot 
    \left( 
    V_{\rho} \hat{\rho} + V_n \hat{n} + V_3 \hat{e}_3^{\prime}
    \right) dt
    = \int_0^{\Delta t_1} \left(
    f_{\rho} V_{\rho} + f_n V_n
    \right) dt
    \\ &
    = \int_0^{\Delta t_1} \left(
    M V_{\rho} \frac{dV_{\rho}}{dt} - \cancel{M \dot{\phi}V_{\rho}V_n} 
    + M V_n \frac{dV_n}{dt} + \cancel{M \dot{\phi} V_nV_{\rho}}
    \right) dt
    = \frac{M V_{h,1}^2}{2} - \frac{M V_{h,0}^2}{2}
\end{aligned}
\end{equation}
where $V_h = \sqrt{V_{\rho}^2 + V_n^2}$ represents the horizontal component of the 
CM velocity. 
The component of $\vec{V}$ along $\hat{e}_3^{\prime}$, $V_3$, does not contribute to 
the calculation of the term in expression \eqref{EnergiaResidual2}. 

Assuming that the cylinder rotates around the contact point without sliding for most 
of the collision time, the components of $\vec{V}$ that form $V_h$ will be given by: 
$V_{\rho} = \omega_n h$ and $V_n = -(\omega_{\rho}h + \omega_3^{\prime} \rho)$. It 
can be observed that $V_{\rho} \propto \omega_n$, which means that as the cylinder 
collides and loses angular velocity around the nodal axis (which is what can cause 
the transition from a ``side'' to a ``face'' state and vice versa), it stops moving 
in the $\hat{\rho}$ direction. On the other hand, $V_n$ consists of two terms, one 
proportional to $\omega_{\rho}$ and the other proportional to $\omega_3^{\prime} = 
\dot{\phi}$. These angular velocity components can be written as:

\begin{equation}
    \begin{aligned}
    \omega_n &= \vec{\omega}^{\prime}\cdot \hat{n} = \omega_1^{\prime} \cos \phi + \omega_2^{\prime} \sin \phi
    \\
    \omega_{\rho} &= \vec{\omega}^{\prime}\cdot \hat{\rho} = - \omega_1^{\prime} \sin \phi + \omega_2^{\prime} \cos \phi
    \end{aligned}
\end{equation}
which is a system that can be solved for $\omega_1^{\prime}$ and $\omega_2^{\prime}$, yielding:

\begin{equation}
    \begin{aligned}
    \omega_1^{\prime} &= - \omega_{\rho} \sin \phi + \omega_n \cos \phi
    \\
     \omega_2^{\prime} &= \omega_{\rho} \cos \phi + \omega_n \sin \phi
    \end{aligned}
\end{equation}

so that when the die is no longer toppling ($\omega_n \rightarrow 0$), the die will 
still be rolling with $\omega_3 = \omega_{\rho}$.


Substituting the results \eqref{EnegiaResidual0}, \eqref{EnergiaResidual1}, and 
\eqref{EnergiaResidual2} into \eqref{PenultimoTermo}, we have:

\begin{equation}
    \begin{aligned}
    W_1 
    &=  \frac{MV_{3,1}^2}{2} + \frac{MV_{h,1}^2}{2} + Mg h_1 - \frac{MV_{3,1}^2}{2} - \frac{MV_{h,0}^2}{2} - Mg h_0
    \\
    & \quad \qquad 
    + \frac{I \omega_{1,1}^{2}}{2} + \frac{I \omega_{2,1}^2}{2} + \frac{I_3 \omega_{3,1}^2}{2}
    - \frac{I \omega_{1,0}^2}{2} - \frac{I \omega_{2,0}^2}{2} - \frac{I_3 \omega_{3,0}^2}{2}
    \end{aligned}
\end{equation}

After the first collision, the energy will be $E_1 = E_0 + W_1$, and after 
successive collisions, we have:

\begin{equation}
    \label{EnergiaColisaoN}
    E_n = E_0 + W_1 + W_2 + \ldots + W_n 
    = K + M g h_n
\end{equation}

where the various dissipative terms cancel out terms from previous terms and convert 
kinetic energy into other forms of energy. In the end, only the potential energy 
associated with the height of the CM remains, plus an additional term $K$ that will 
be zero for the ``face'' states and equivalent to a kinetic energy 
$\frac{1}{2}\left( M V^2 + I_3 \omega_3^2 \right)$ for the ``side'' state.

\subsection{The statistics of the results of cylinder tosses}

It is obvious, at this point, that it would be a terrible idea to try to analyze the 
motion, even of a single die, from a deterministic perspective. Small perturbations 
in the initial conditions and the characteristics of the point of contact with the 
surface of the landing plane would already result in significant differences in the 
sequence of movements. The explanations in the previous sections, therefore, do not 
intend to follow that path but rather to glimpse the characteristics of the motion, 
infer qualitatively what to expect from it, and finally, understand how the 
definition of a ``face'' or ``side'' state will be associated with a certain amount 
of energy that remains after a toss and the subsequent collisions that follow.

Hence, given the dimensions of a given cylinder, it can end up in a ``side'' state 
with energies $E_S(V)$ or ``face'' state with energy $E_F$. Thus, a die is a two-
level system.

If dice are placed on a plane that ejects them, constantly shaking and causing a 
sequence of random collisions, we can say that these dice receive an average energy 
from this plane, which is partly converted into potential energy, propelling the 
dice upward, partly into translational kinetic energy, and partly into rotational 
kinetic energy. Thus, the dice have a certain probability of receiving energy from 
the landing/launching plane and then retaining part of that energy according to the 
discussion in the previous section. If we think of a large number of identical 
copies of this system that launch dice and where dice have a certain probability of 
receiving energy, the comparison with a canonical ensemble where multiple systems 
are in contact with a certain thermal reservoir at temperature $T$ is inevitable. A 
canonical ensemble, in turn, follows a Boltzmann probability distribution 
function~\cite{Blundell}.

It is evident, however, that this idea, although aesthetically appealing, is 
limited, especially because it would be impractical to launch a large number of 
cylindrical dice on the order of $1 \, \mathrm{mol}$. Nevertheless, we will work 
with this hypothesis.

\subsubsection{Statistics of free fall}

We can start by revisiting Equation \ref{EnergiaMecanica}, which describes the 
energy of a cylinder during free fall. Knowing this, we can calculate the partition 
function and, subsequently, the average energy of this system as a function of 
$\beta = 1/(k_b T)$. For this purpose, we define $Z_{pot}$, $Z_{kin}$, and 
$Z_{rot}$, which, when multiplied, result in the desired partition function $Z$.

\begin{equation}
    \label{free fall partition function}
    \begin{aligned}
        Z_{pot} &= \int\limits_{0}^{\infty} e^{- \beta M g h}\, dh
        \\ 
        Z_{kin} &= \left(\int\limits_{-\infty}^{\infty} e^{- \beta \frac{M V^{2}}{2}}\, dV\right)^{3}
        \\
        Z_{rot} &= \left(\int\limits_{-\infty}^{\infty} e^{- \beta \frac{I\omega^{2}}{2}}\, d\omega\right)^{2} \int\limits_{-\infty}^{\infty} e^{- \beta \frac{I_{3} \omega_{3}^{2}}{2}}\, d\omega_{3}
    \end{aligned}  
\end{equation}

The value of $Z_{pot}$ introduces the gravitational potential energy. The values of 
$Z_{kin}$ and $Z_{rot}$ introduce, each, 3 degrees of freedom related to 
translational and rotational kinetic energies, respectively. After performing the 
integrals, we can calculate $Z$:

\begin{equation}
    \label{free fall Z}
    Z =  \frac{8 \pi^{3}}{I \sqrt{I_{3}} \beta^{4} g M^{\frac{5}{2}}}
\end{equation}

The average energy is given by:

\begin{equation}
    \label{free fall average energy}
    \left< E \right> = - \frac{1}{Z} \frac{dZ}{d \beta}=  \frac{4}{\beta}
\end{equation}

By isolating $\beta$, it is possible to find this value as a function of the average 
energy of the system. However, this energy depends on time since there is 
dissipation due to collisions. Thus, we have:

\begin{equation}
    \label{beta value}
    \beta \left( t \right) = \frac{4}{\left< E\left( t \right) \right>}
\end{equation}

Note that in $Z_{pot}$, the height $h$ was integrated with a lower limit of $0$. 
This means that the reference for gravitational potential energy was shifted from 
the ground to the minimum height, making the involved calculations simpler. 
Therefore, this energy present in the $\beta$ formula is actually an energy 
variation, i.e., the energy received by the plane minus the dissipated energy in 
collisions.

\subsubsection{Statistics of the final state}

Having done that, it is necessary to consider the final state. Unlike free fall, the 
cylinder is forced to assume one of the two previously mentioned states: ``face'' or 
``side''. Knowing the energies of each state, we can write a new Boltzmann 
distribution, with different partition function and probabilities from the previous 
section.

The energy of the ``side'' state is given by:

\begin{equation}
    \label{E_S}
    E_S = Mgh_S + K = MgR + \frac{3MV^2}{4}
\end{equation}
where the value of kinetic energy $K$ was calculated considering a rolling without 
slipping constraint. However, the kinetic energy must be understood as a second 
possibility, meaning that the cylinder can fall into this state with only potential 
energy or with the presence of kinetic energy.

The energy of the ``face'' state is:
\begin{equation}
    \label{E_F}
    E_F = Mgh_F = \frac{MgH}{2}
\end{equation}

Thus, we can calculate the respective probabilities. Let $P_S$ be the probability of 
observing the die in the ``side'' state, and $q = H/R$:
\begin{equation}
    \label{P_S}
    P_S = \frac{1}{Z} \left( \frac{1}{Z'} \ C_{S} \int\limits_{0}^{\infty} 2 \pi V e^{- \beta \left(M g R + \frac{3 M V^{2}}{4}\right)}\, dV + C_{S} e^{- \beta M g R} \right)
\end{equation}
where $C_S$ is a coefficient related to the multiple ways of finding the ``side'' 
state  with the same energy, and $Z'$ is a proportionality constant, which will be 
both discussed later.

Since a velocity vector of magnitude $V$ can point in different directions in space, 
we also needed to consider these possibilities, and for that, we think in terms of 
velocity space. An area element $dA$ in this space can be calculated as:

\begin{equation}
    \label{velocity space}
    dA = \int_{0}^{2 \pi} \,d\theta \int_{V}^{V + dV} r \,dr = 2\pi V \, dV
\end{equation}
thus, the term $2 \pi V$ present in the expression was explained.

However, upon analyzing the dimension of the term resulting from the integral, it is 
noticed that it has units of velocity squared. As a probability must be 
dimensionless, we added the element $1/Z'$, with $Z'$ having the same unit as the 
term in question, so the dimensions cancel out.

Traditionally, velocity space would be used for comparisons between restricted areas 
or volumes and the total. A clear example is the calculation of the probability of 
finding a particle in a gas within a certain range of velocity magnitudes, by 
comparing a specific volume, represented by a sphere, with respect to the total 
volume of space. It is evident that neither the total volume nor the partial volume 
represents the actual number of microstates, but they are proportional to these 
values, enabling the calculation of probabilities.

In the discussed case, we have to compare a term calculated using velocity space 
with another in which this was not used, which is, at first, incompatible. However, 
we know that the quantity of microstates is proportional to the result of the 
integral, so the sought-after value must be this result multiplied by a constant $1/
Z'$. Although we do not know the value of $Z'$, we can speculate that it represents 
an area in velocity space:

\begin{equation}
    \label{Z'}
     Z' = \pi \, V_{Z'}^2
\end{equation}
in which $V_{Z'}$ can be something like the typical or the maximum velocity reached 
by the cylinders.

The expression for the probability of the ``face'' state is considerably simpler, 
due to the absence of kinetic energy:

\begin{equation}
    \label{P_F}
    P_F = \frac{1}{Z} \ C_F \ e^{-\beta E_F}
\end{equation}
and here, once again, $C_F$ is a coefficient related to the multiple ways of finding 
the ``face'' state with the same energy.

Having done that, we are in a position to solve the integral of $P_S$, calculate the 
partition function by normalization, and write the probabilities. However, the 
coefficients of multiplicity to be calculated still remain.
Hence, the expression for $P_S$ becomes:

\begin{equation}
    \label{P_L final}
    P_S = \frac{C_{S} e^{- \beta M g R} + \frac{1}{Z'} \frac{4 \pi}{3 \beta M} C_{S} e^{- \beta M g R} }{C_{F} e^{- \beta \frac{M g R q}{2}} + C_{S} e^{- \beta M g R} + \frac{1}{Z'} \frac{4 \pi }{3 \beta M} C_{S} e^{- \beta M g R} }
\end{equation}

\subsubsection{Calculation of $C_S$ and $C_F$}

Before we proceed with the calculation itself, let's elaborate a bit more on the 
need for these coefficients. Imagine that, instead of cylinders, the solids in 
question were pyramids. Intuitively, we know that it is impossible for such a 
pyramid to come to rest balanced on its vertex without piercing the surface or being 
glued to it. This is due to the lack of stability of this state, which would quickly 
transform into a state balanced on one of the faces of the pyramid. However, if we 
were to write the probability of this state, it would be extremely higher than what 
is observed experimentally because the multiple ways of finding each state with the 
same energy were not taken into account. To overcome this problem, we introduce 
these coefficients.

\begin{figure}[h!]
    \begin{center}
        \includegraphics[width=0.4\textwidth]{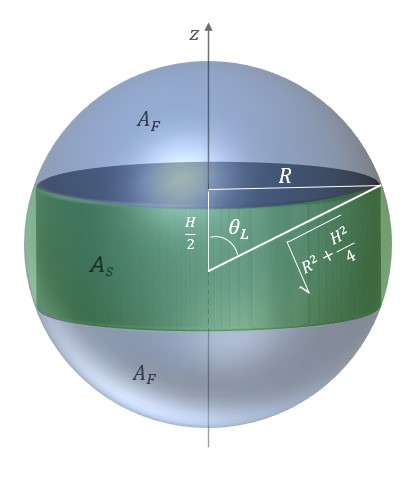}
    \end{center}
    \caption{Sphere circumscribed around the cylinder.
    \label{esfera}
    }
\end{figure}

For the calculation of these values, we will consider a sphere circumscribed around 
the cylinder (see Figure \ref{esfera}). Imagine a cylinder in free fall, but in the 
reference frame of the object itself. In this case, the ground would be approaching, 
and all the different ways that could happen would form a sphere around the 
cylinder. Certain parts of the sphere are associated with a collision on the face 
part, $A_F$, and others on the side part, $A_S$. Thus, we will propose that the area 
corresponding to each part, divided by the total area of the sphere, is equivalent 
to the coefficient of the respective state.

\begin{equation}
    \label{A_F}
    A_F = 2\int_{0}^{2 \pi} \,d\phi \int_{0}^{\theta_L} R^2 \sin{\theta}\,d\theta 
\end{equation}

\begin{equation}
    \label{A_L}
    A_S = 4 \pi R^2 -A_F
\end{equation}

Thus, we can write the coefficients:
\begin{equation}
    \label{C_S}
    C_S = \frac{A_L}{4 \pi R^2} = \left( \frac{q}{\sqrt{q^2 + 4}} \right)
\end{equation}

\begin{equation}
    \label{C_F}
    C_F = \frac{A_F}{4 \pi R^2} =   \left( 1- \frac{q}{\sqrt{q^2 + 4}} \right)  
\end{equation}

Note that these values are normalized, i.e., they range from 0 to 1 and sum up to 1. 
Mathematically, $C_F + C_S = 1$.

\subsubsection{Estimating $\beta$}
\label{estimation}

According to Equation \ref{beta value}, the final value of $\beta$ occurs when $t 
\rightarrow \infty$. To calculate this value, we need to know the final average 
energy, which can be written as follows:

\begin{equation}
    \label{final average energy}
    \left< E_f \right> = \left<E_0\right> + \left<W\right>
\end{equation}
where $\left<E_0\right>$ is the average energy received from the plane and 
$\left<W\right>$ is the sum of the energies dissipated after successive collisions.

In order to estimate the value of $\left<W\right>$, we will propose that there are 
two types of collisions. The first type is related to a collision in the $A_F$ 
region, while the second type is related to a collision in the $A_S$ region. Thus, 
we will also assume the existence of two types of work, related to a series of 
collisions in each of these areas. If $\left<W\right> = W_F$, only collisions in 
$A_F$ have occurred. On the other hand, if $\left<W\right> = W_S$, only collisions 
in $A_S$ have occurred.

It is clear that these two cases are unreal, and a combination of these two types of 
collisions is expected. The coefficients of multiplicity, calculated earlier, can 
indicate the contribution expected from each type of work. For example, if 
$C_S = 0.7$ and $C_F = 0.3$, it is expected that 7 out of 10 collisions have 
occurred in the $A_S$ region. Similarly, we can use the same relationships for the 
works:

\begin{equation}
    \label{average W}
    \left<W\right> = C_F W_F + C_S W_S  
\end{equation}

However, it is obvious that this proposal is only an approximation. It is evident 
that there is variation in the values of $W_S$ and $W_F$ even if the collisions are 
in their respective regions, i.e., depending on how this collision occurs, there 
will be more or less dissipation. This estimation method will be more efficient in 
cases where a higher collision rate is not forced in any of the regions, which is 
valid for the experiment of the plane ejecting cylinders. However, if the cylinders 
are horizontally launched and always in the same way, there will be a greater 
tendency for collision in a specific manner, causing the coefficients to have values 
that diverge from what is observed, requiring a correction.

Furthermore, it is important to note that the absolute value of $W_F$ should be 
greater than the absolute value of $W_S$. This occurs because the ``side'' state 
allows for rolling, dissipating less energy as it remains in the form of rotation.

\section{\label{sec:descricao}Experimental Description}

\subsection{\label{subsec:equipment}Used Equipment and Experimental Setup}

As described in the theoretical fundamentals section, the launch conditions of the 
cylinder greatly affect $P_S$ and $P_F$. Therefore, it is necessary to develop an 
experiment to standardize each repetition. Considering the need for a large data 
sample, the solution found was the construction of a machine using Arduino and, 
subsequently, a program in the Python programming language to recognize the two 
possible states, ``face'' and ``side'', and calculate their respective probabilities.

A box made of medium-density fiberboard (MDF) was used to house the entire 
experiment. The base, with a thickness of $6 \,\mathrm{mm}$, has a cavity with a 
depth of $3 \,\mathrm{mm}$, where the protoboard (a board used for circuit assembly) 
was inserted. The side walls of the box were made of the same material as the base 
and half the thickness, i.e., $3 \,\mathrm{mm}$. Structures for support and 
connection between the walls were 3D printed using polylactic acid (PLA). Two 
cardboard sheets, one covered with sulfite paper and the other with suede, were also 
prepared as surfaces for launching the cylinders.

The cylinders used were 3D printed in white PLA with a $25\%$ infill, and each face 
of the same cylinder was painted blue and red, respectively, for recognition 
purposes. The radius was kept constant at $7.5 \,\mathrm{mm}$, and the $H/R$ ratios 
varied from $0.3$ to $2.5$ with intervals of $0.1$.

\begin{figure}[h!]
    \begin{center}
        \includegraphics[width=0.3\textwidth]{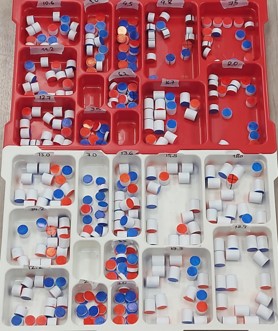}
    \end{center}
    \caption{Cylinders used in the experiments.
    \label{cilindros}
    }
\end{figure}

\begin{figure}[h!]
    \begin{center}
        \includegraphics[width=0.85\textwidth]{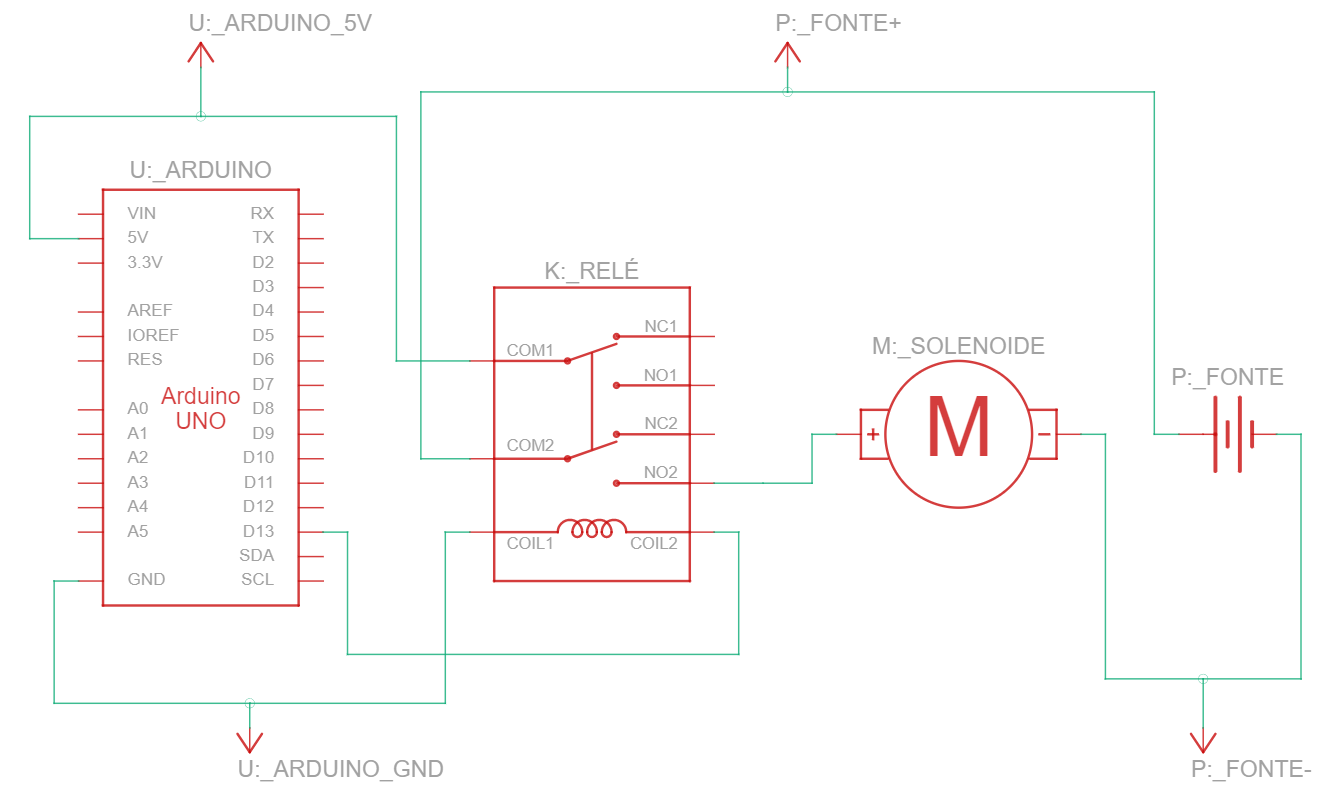}
    \end{center}
    \caption{Electrical schematic of the system for a single solenoid.
    \label{circuito}
    }
\end{figure}

For the assembly of the electrical system (Figure~\ref{circuito}), an Arduino UNO 
board was used, as well as 8 JF-0530B model solenoids, each with a force of $5 \,
\mathrm{N}$, controlled by relays (electromechanical switches). To provide the 
necessary voltage for the motor operation, a regulated DC power supply with a 
voltage of $22 \,\mathrm{volts}$ was used. For image capture, a Logitech C920 camera 
positioned above the box, fixed on a universal stand, was controlled by a computer 
program.

\begin{figure}[h!]
    \begin{center}
    \mbox{
    \subfigure[]{\includegraphics[width=0.4\columnwidth]{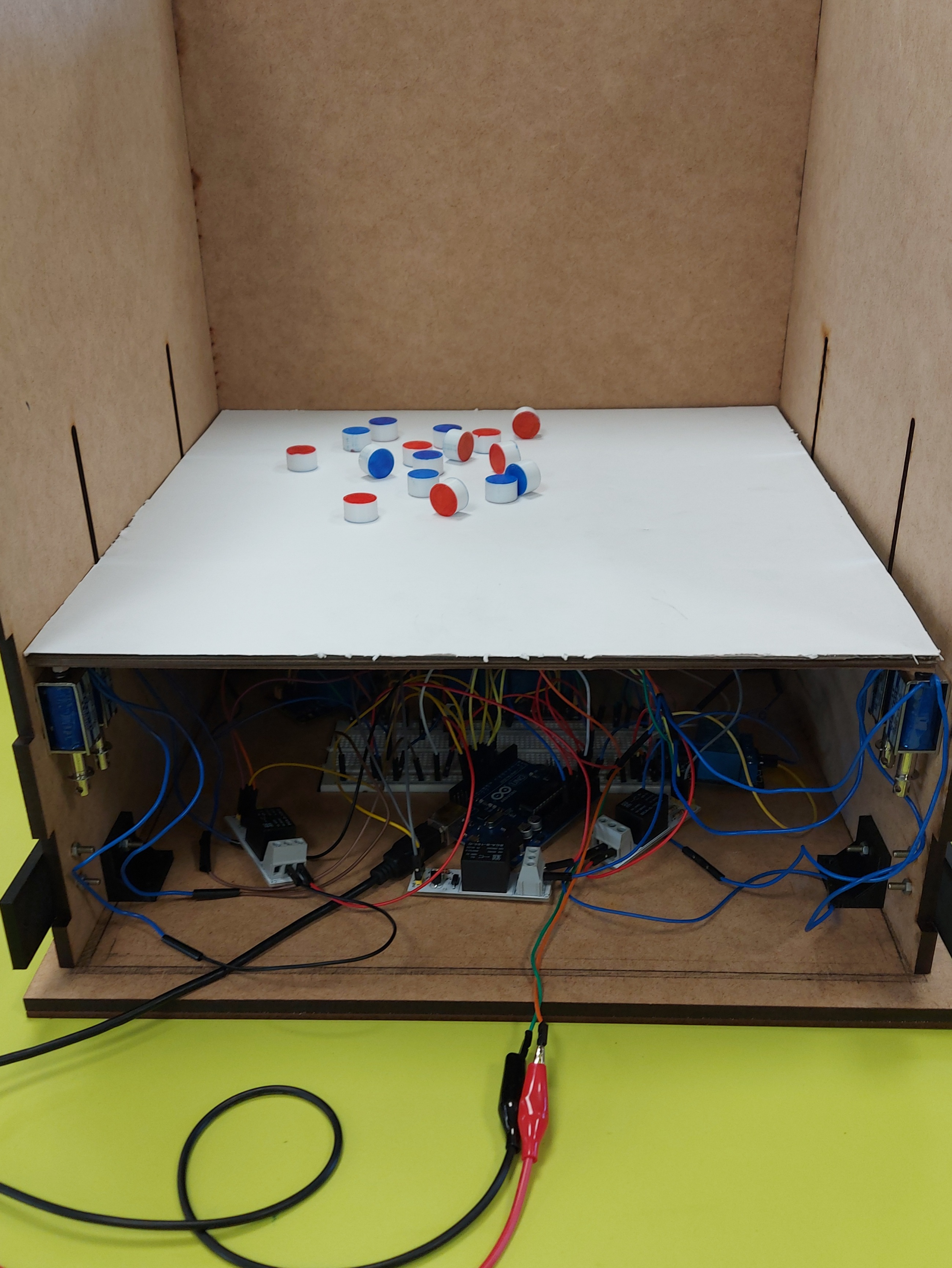}\label{maquinaA}}
    \subfigure[]{\includegraphics[width=0.4\columnwidth]{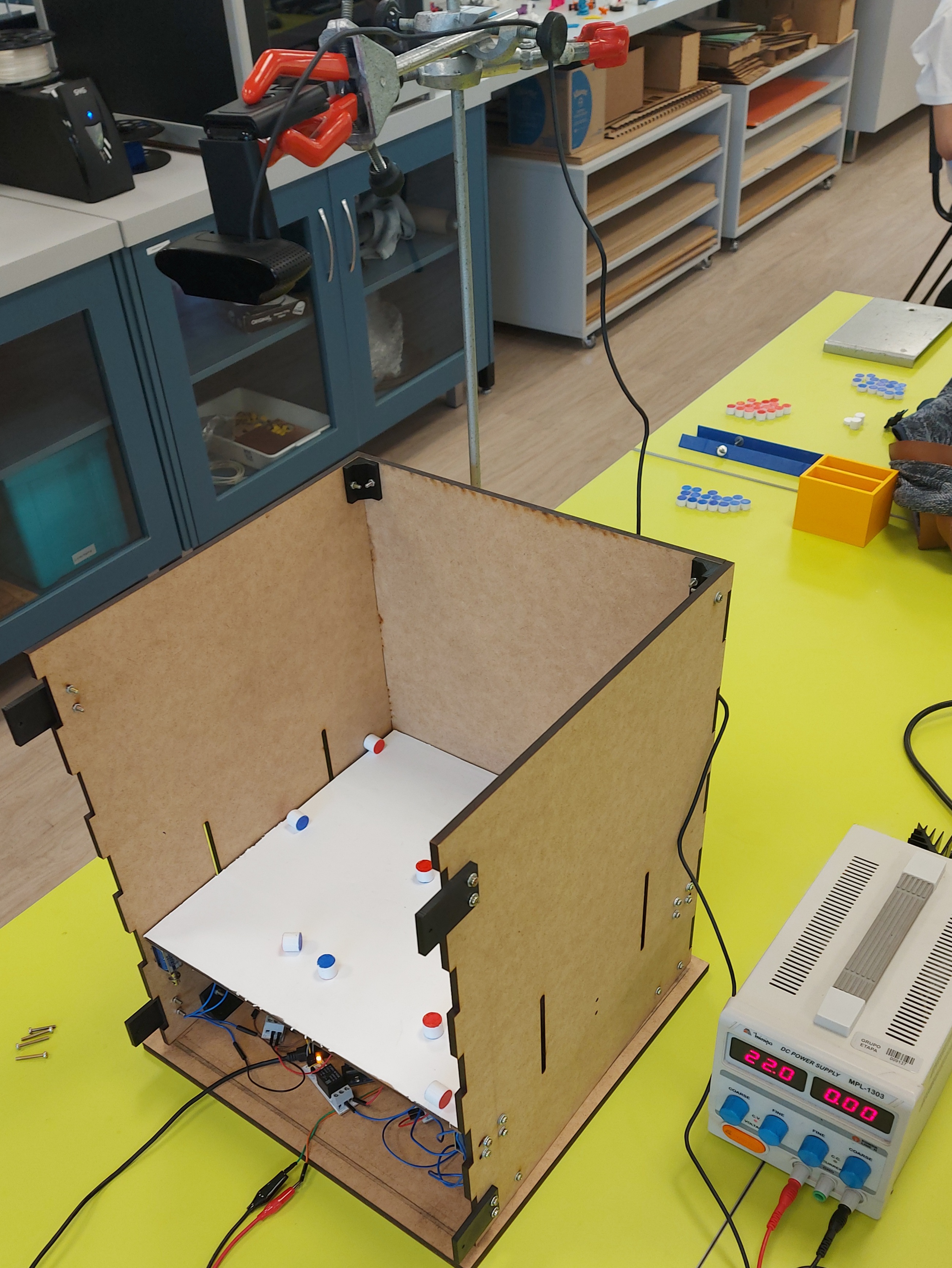}\label{maquinaB}}
    }
    \end{center}
    \caption{(a) Front view of the system; (b) Complete view of the system with the camera.
    \label{maquina}
    }
\end{figure}

With all the mentioned components, the system was assembled (Figure~\ref{maquina}), 
with the solenoids reaching the cardboard plate, lifting it and performing a launch.

Additionally, for the determination of some parameters, namely the coefficient of 
restitution ($\varepsilon$), the average height reached by the cylinders 
($\bar{h}$), and the static and kinetic friction coefficients ($\mu_s$ and $\mu_k$), 
the system was modified. For the first two parameters, the system was set up without 
the front wall, and the camera was repositioned to capture the movement from the 
front (Figure~\ref{SistemaAltura}). For the friction coefficients, the sulfite-
coated cardboard plate was placed on an inclined plane, and the sliding of a 
cylinder was analyzed.

\subsection{\label{subsec:procedure}Procedure}

\begin{figure}[h!]
    \begin{center}
    \mbox{
    \subfigure[]{\includegraphics[width=0.45\columnwidth]{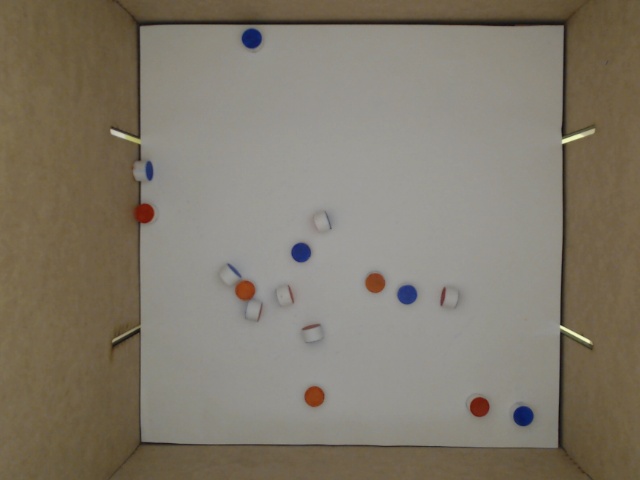}\label{FotoCilindros}}
    \subfigure[]{\includegraphics[width=0.45\columnwidth]{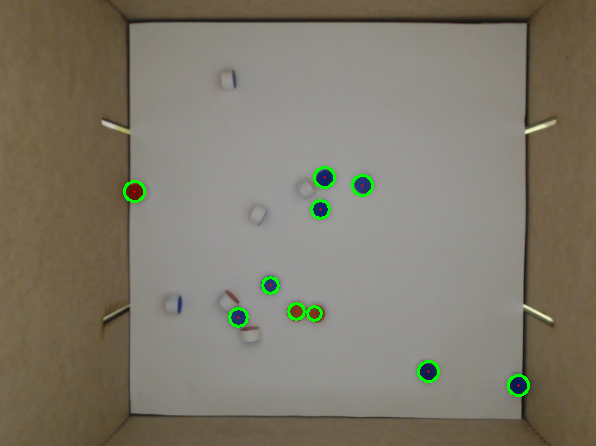}\label{FotoReconhecimento}}
    }
    \end{center}
    \caption{(a) Image captured by the camera; (b) Circle recognition by the algorithm.
    \label{fotos1}
    }
\end{figure}

With the system assembled, a Python program was executed to control the Arduino and, 
consequently, the movement of the solenoids through serial signals. The solenoids 
hit the cardboard plate on which the cylinders were placed, launching them. After 
the launch, the program waited for 5 seconds to allow the cylinders to stabilize and 
then activated the camera to capture an image (Figure~\ref{FotoCilindros}) of the 
cylinders in their respective final states (``face'' or ``side'').

This process was repeated two hundred times for each $H/R$ ratio (ranging from $0.3$ 
to $2.5$).

Through this procedure, it was possible to automatically obtain hundreds of photos 
per hour, thus obtaining a large sample. To analyze the final state of each 
cylinder, a second algorithm was programmed to recognize circles and colors 
(Figure~\ref{FotoReconhecimento}) and identify whether they were in the ``face'' or 
``side'' state.

Knowing the number of times the final result was ``face'' or ``side'' and the total 
number of launches, the probability of each state was calculated, and experimental 
graphs of $P_L$ and $P_F$ as a function of the $H/R$ ratio (which can also be 
expressed solely as a function of H since R was kept constant) were created using 
these data points.

\begin{figure}[h!]
    \begin{center}
    \mbox{
    \subfigure[]{\includegraphics[width=0.45\columnwidth]{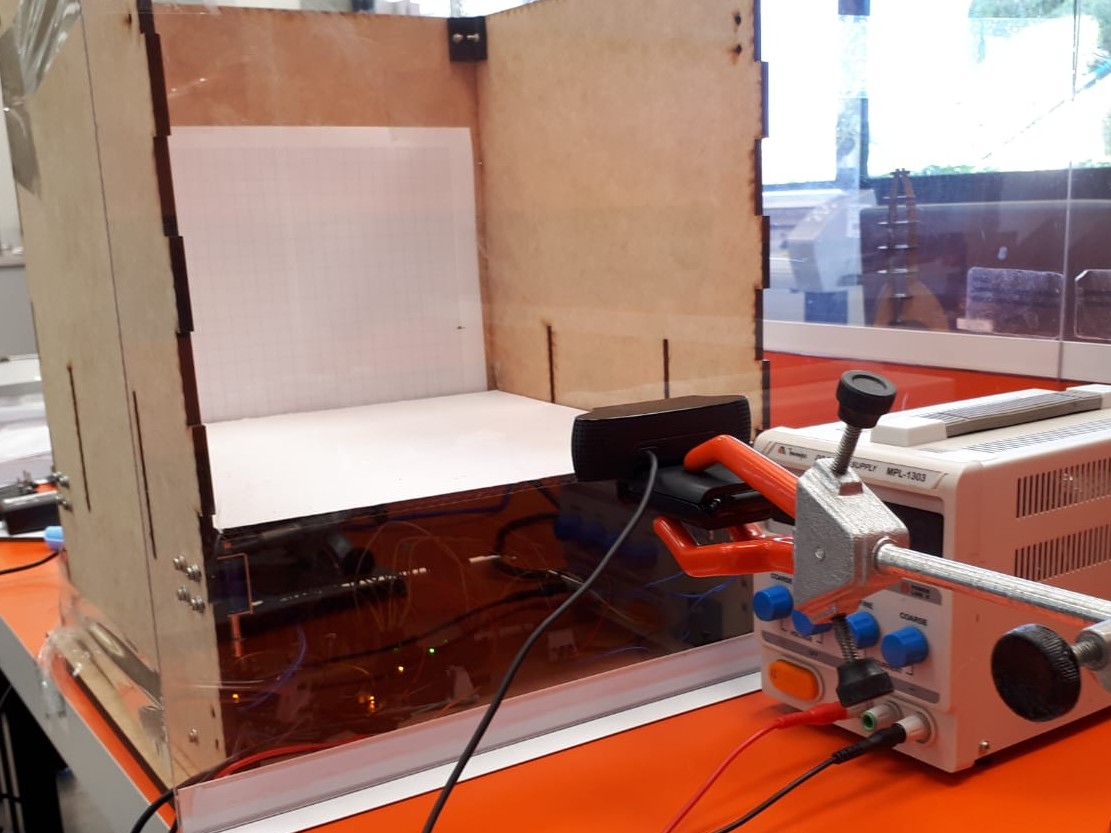}\label{SistemaAltura}}
    \subfigure[]{\includegraphics[width=0.45\columnwidth]{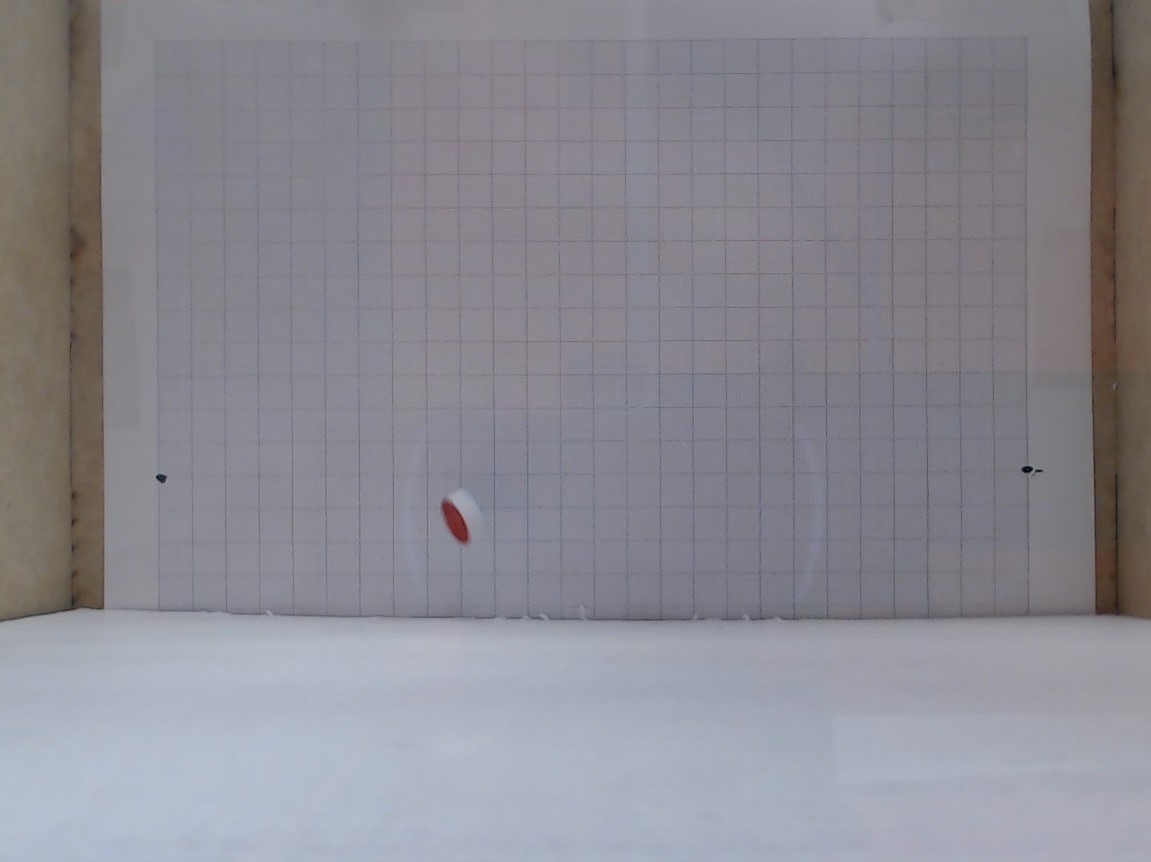}\label{AlturaCilindro}}
    }
    \end{center}
    \caption{(a) Image of the system with the front-facing camera positioned; (b) Image captured by the camera with the position of the grid paper.
    \label{fotos2}
    }
\end{figure}

By modifying the system for the setup with the front-facing camera 
(Figure~\ref{SistemaAltura}), the average height ($\bar{h}$) and the coefficient of 
restitution ($\varepsilon$) were determined. A millimeter grid paper was used on the 
back wall of the box (Figure~\ref{AlturaCilindro}) to enable measurement.

Simultaneously, experiments were conducted to determine the friction coefficients. 
For the static coefficient, the plate was placed on the plane without inclination, 
and the angle was gradually increased until the sliding threshold was reached. For 
the kinetic coefficient, the plate was inclined above the maximum angle of static 
friction, and a cylinder was released, with the time of motion measured.

\subsection{Secondary Experiment: Horizontal Launch}

Although the developed theory is much more adapted to the machine case, we also 
created a second experiment to test the limits of this formulation. For a horizontal 
launch, there is a significant increase in velocity in that direction, inducing 
collisions in a specific region. This situation falls into what was discussed in the 
estimation of $\beta$ (Section \ref{estimation}) and requires a correction in the 
multiplicity coefficients.

For this experimental setup, equipment similar to a catapult was used, which 
operates based on a counterweight. In this way, it is possible to ensure the same 
initial energy for all cylinders, which will be horizontally launched.

\begin{figure}[h!]
    \begin{center}
        \includegraphics[width=0.5\textwidth]{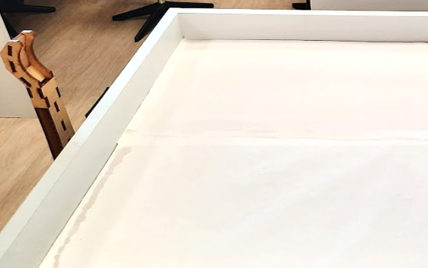}
    \end{center}
    \caption{Image of the experimental setup for horizontal launch.
    \label{catapulta}
    }
\end{figure}



\section{\label{sec:analise}Discussion and Data Analysis}

\subsection{Obtained Results}

To analyze the probabilities of each state for each $H/R$ ratio, the first step was 
to determine $\bar{h}$. The experimentally obtained result was: $\bar{h} = 3.8\,
\mathrm{cm}$ for $q = 1$. With this value, it is possible to calculate the energy 
supplied by the plane to the cylinders.

The second step would be to calculate the values of $W_S$ and $W_F$. However, the 
calculation of these numbers, both theoretically and experimentally, remains as a 
topic for future research. Thus, an adjustment was made for these values, and for 
better understanding, they were normalized by dividing them by $E_0$. To perform the 
adjustment, the computer program starts with both values set to $0.5$ and adjusts 
them in small increments until the standard deviation of the theoretical and 
experimental values is minimized.

Starting with the results obtained with sulfite paper, the fit resulted in the 
values given in \eqref{fit-sulfite-1}.

\begin{equation}
    \label{fit-sulfite-1}
    \begin{aligned}
        W_S &= 0.475 \ \pm \ 0.001
        \\
        W_F &= 0.999 \ \pm \ 0.001
    \end{aligned}  
\end{equation}
from which we have graphs of $W$ and $\beta$ as functions of $q$.

%

\begin{figure}[h!]
    \begin{center}
        \mbox{ 
            \subfigure[]{
                \includegraphics[width=0.48\columnwidth]{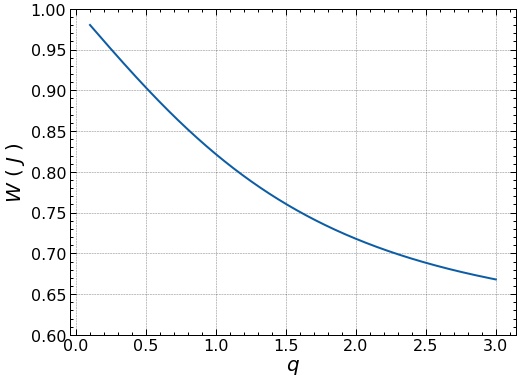}
                \label{cartolina-W}
                }
            \subfigure[]{
                \includegraphics[width=0.48\columnwidth]{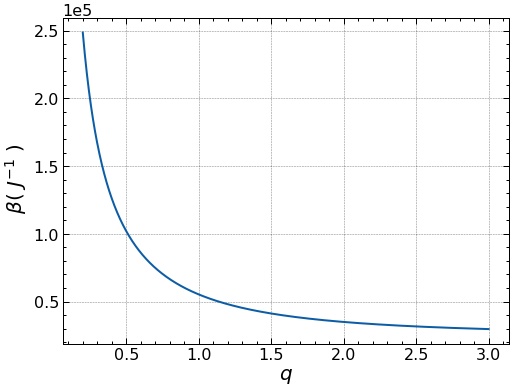}
                \label{cartolina-beta}
                }
        }
    \end{center}
    \caption{Graphics for sulfite paper in the machine displaying (a) $W$, and (b) $\beta$ as  functions of $q$.
    \label{cartolina-maquina}
    }
\end{figure}

Moving on to the results obtained with suede fabric, the fit resulted in the 
values given in \eqref{fit-camurca}.

\begin{equation}
    \label{fit-camurca}
    \begin{aligned}
        W_S &= 0.836 \ \pm \ 0.001
        \\
        W_F &= 0.878 \ \pm \ 0.002
    \end{aligned}  
\end{equation}

%

\begin{figure}[h!]
    \begin{center}
        \mbox{ 
            \subfigure[]{
                \includegraphics[width=0.48\columnwidth]{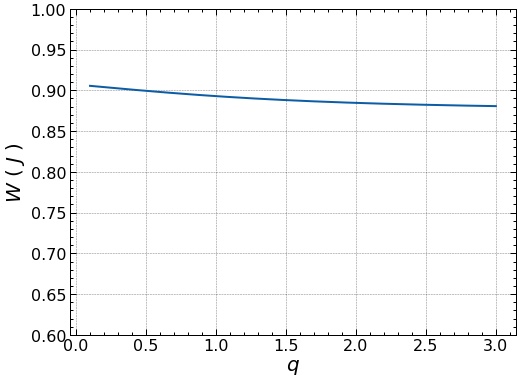}
                \label{camurca-W}
                }
            \subfigure[]{
                \includegraphics[width=0.48\columnwidth]{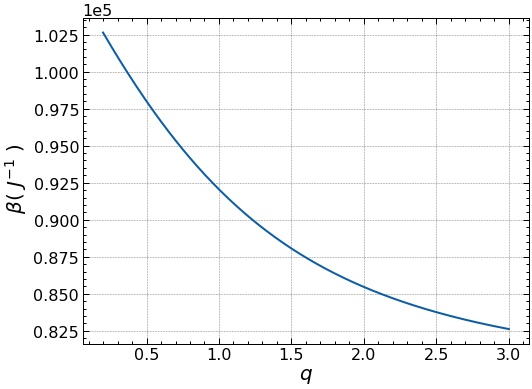}
                \label{camurca-beta}
                }
        }
    \end{center}
    \caption{Graphics for suede fabric in the machine displaying (a) $W$, and (b) $\beta$ as  functions of $q$.
    \label{camurca-maquina}
    }
\end{figure}

\begin{figure}[h!]
    \begin{center}
        \mbox{ 
            \subfigure[]{
                \includegraphics[width=0.48\columnwidth]{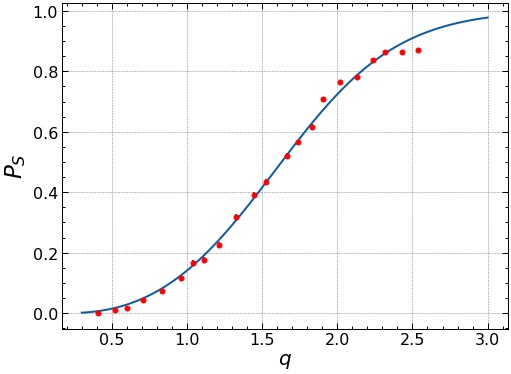}
                \label{cartolina-pl}
                }
            \subfigure[]{
                \includegraphics[width=0.48\columnwidth]{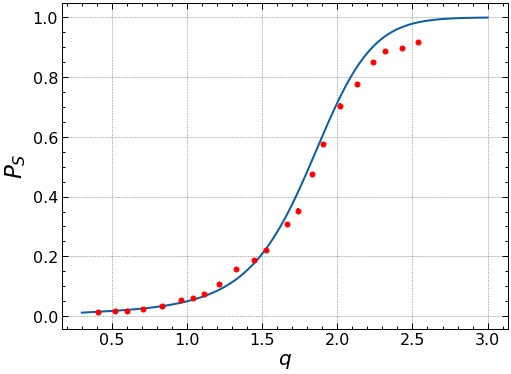}
                \label{camurca-pl}
                }
        }
    \end{center}
    \caption{Graphics of $P_S$ as a function of $q$ in the machine for (a) sulfite paper, and (b) suede fabric.
    \label{pl-maquina}
    }
\end{figure}

Finally, in Figure~\ref{pl-maquina}, we have the graphs of $P_S$ as a function of 
$q$ being compared with the experimental results obtained.

%

\subsection{Horizontal Launch}

With the help of the Tracker software, we can calculate the velocity and height at 
which the cylinders are ejected and, therefore, the energy supplied. The fits for 
the values of $W_S$ and $W_F$ are as follows for each surface.

\begin{figure}[h!]
    \begin{center}
        \mbox{ 
            \subfigure[]{
                \includegraphics[width=0.48\columnwidth]{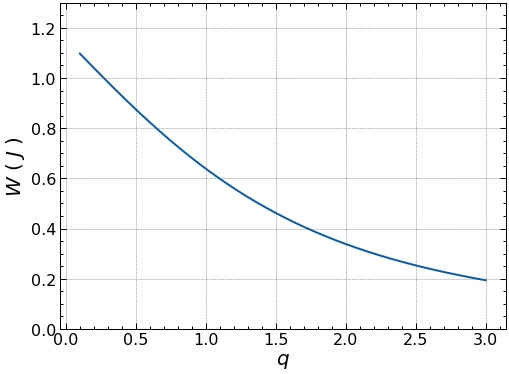}
                \label{cartolina-hor-W}
                }
            \subfigure[]{
                \includegraphics[width=0.48\columnwidth]{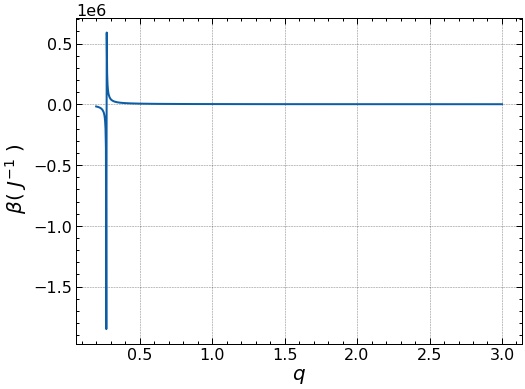}
                \label{cartolina-hor-beta}
                }
        }
    \end{center}
    \caption{Graphics for sulfite paper in the horizontal launch displaying (a) $W$, and (b) $\beta$ as  functions of $q$.
    \label{cartolina-horizontal}
    }
\end{figure}

For sulfite paper, we found the values given in \eqref{fit-sulfite-2}. 

\begin{equation}
    \label{fit-sulfite-2}
    \begin{aligned}
        W_S &= 0.00 \ \pm \ 0.01
        \\
        W_F &= 1.142 \ \pm \ 0.001
    \end{aligned}  
\end{equation}
and, as we did before for the machine, we produced the resulting graphs for these 
values, which can be seen in Figure~\ref{cartolina-horizontal}.

%

\begin{figure}[h!]
    \begin{center}
        \mbox{ 
            \subfigure[]{
                \includegraphics[width=0.48\columnwidth]{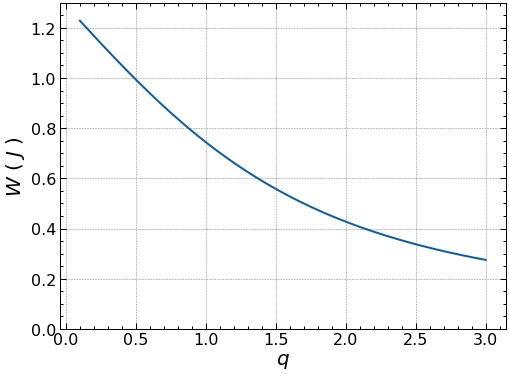}
                \label{camurca-hor-W}
                }
            \subfigure[]{
                \includegraphics[width=0.48\columnwidth]{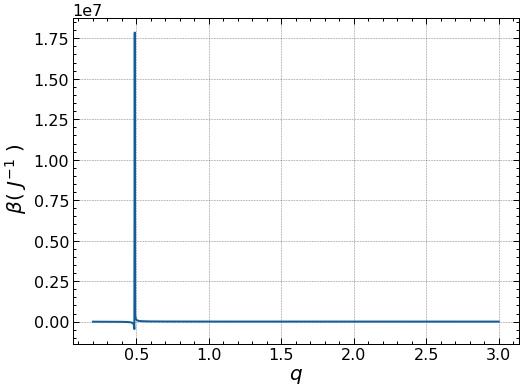}
                \label{camurca-hor-beta}
                }
        }
    \end{center}
    \caption{Graphics for suede fabric in the horizontal launch displaying (a) $W$, and (b) $\beta$ as  functions of $q$.
    \label{camurca-horizontal}
    }
\end{figure}

For suede fabric, the values found were those given in \eqref{fit-camurca-2}.

\begin{equation}
    \label{fit-camurca-2}
    \begin{aligned}
        W_S &= 0.000 \ \pm \ 0.005
        \\
        W_F &= 1.3118 \ \pm \ 0.0005
    \end{aligned}  
\end{equation}
with resulting graphs shown in Figure~\ref{camurca-horizontal}.

%

\begin{figure}[h!]
    \begin{center}
        \mbox{ 
            \subfigure[]{
                \includegraphics[width=0.48\columnwidth]{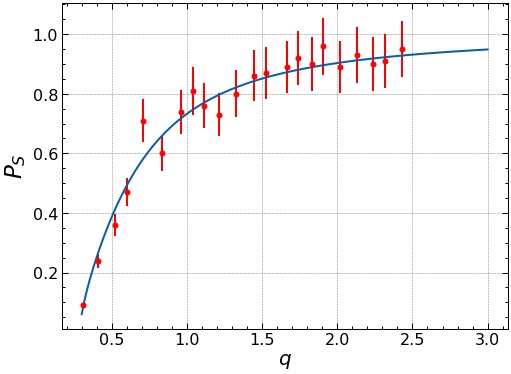}
                \label{cartolina-hor-pl}
                }
            \subfigure[]{
              \includegraphics[width=0.48\columnwidth]{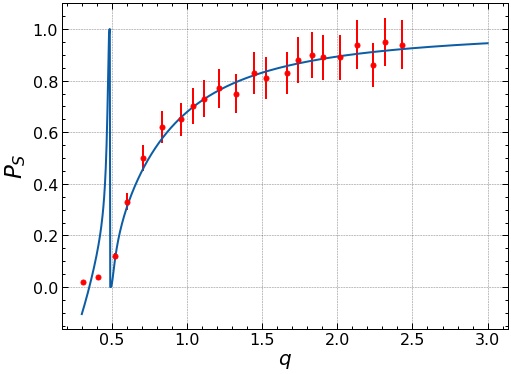}
                \label{camurca-hor-pl}
                }
        }
    \end{center}
    \caption{Graphics of $P_S$ as a function of $q$ in the horizontal launch for (a) sulfite paper, and (b) suede fabric.
    \label{pl-horizontal}
    }
\end{figure}

The respective results for $P_S$ as a function of $q$ can be seen in
Figure~\ref{pl-horizontal}.

%

\subsection{Discussion of Results}

Analyzing the results of the experimentally obtained probabilities and adjusting the 
theoretical curves, it can be observed that the data, in general, behave as expected 
for the machine. However, there are deviations, as expected, considering that 200 
launches were performed for each $H/R$ ratio, with 16 cylinders in each launch, 
resulting in a total of 3200 cylinders launched for each ratio, which is still a 
small number compared to $1\,\mathrm{mol}$.

These deviations are mainly due to experimental errors, limitations, and theoretical 
approximations. Possible sources of experimental errors include problems with 
computational recognition or, perhaps, wear and deformation of materials due to 
repeated use. Additionally, the experiment with the suede material was conducted 
after the one with the sulfite paper, which means that the materials were much more 
worn, and it is precisely in this experiment that the largest divergences from the 
expected results occurred. As for the limitations in theory, they are due to certain 
factors that were not considered in the calculation of $E$ and, of course, the fact 
that the number of data points is small compared to the statistical limit 
($N \sim 1 \,\mathrm{mol}$).

Thus, by analyzing the behavior of the experimental data and the adjusted curve, it 
is possible to determine the ideal $H/R$ ratio for specific conditions, such as 
materials and launching methods.

For the secondary experiment, the discrepancies are significant for two main 
reasons. Firstly, only 200 launches were performed for each ratio, which is a much 
smaller number of repetitions compared to the machine experiment and further away 
from $1\,\mathrm{mol}$. Secondly, this launching method favors certain forms of 
collision, which means that the calculated multiplicity coefficients are not 
appropriate. This is reflected in the fact that the values of $W_F$ are greater than 
1, which should not occur since the dissipation cannot be greater than the initial 
energy. However, this unexpected value exists to compensate for a very low value of 
$C_F$. Despite these anomalies, which generate peaks in the graphs, the 
probabilities reasonably correspond to what was theoretically predicted for larger 
ratios.



\section{\label{sec:conclusao}Conclusions}

Therefore, it can be concluded that the developed theory provides a reasonable 
approximation for the probabilities of final states of a cylinder based on its $H/R$ 
ratio, through experimental adjustments. However, it would be possible to obtain the 
theoretical values of the factors used in the adjusted function by improving the 
experimental conditions and extending the data analysis to consider other aspects. 
This would enable a comparison between the adjusted values and the corresponding 
theoretical curve.

Furthermore, it remains to explore more thoroughly various possible variations of 
the system, such as different surfaces (beyond the two already used), thus modifying 
the coefficients of friction and restitution, which should be taken into account in 
a more detailed future analysis. These factors can likely be considered after 
further development of the theory. Additionally, limitations of the theoretical 
approach through Statistical Mechanics persist.

The horizontal launch clearly demonstrates the limitations of what has been 
developed so far. In future analyses, as a way to complement what has already been 
done, a more detailed calculation of the multiplicity coefficients and the work done 
in collisions is needed.

Despite its limitations, the theory discussed has applications that go beyond a 
specific solid. For example, if a solid has a symmetry such that the energy of each 
state is the same and the multiplicity coefficients are also the same, it is 
possible to affirm that this solid is a ``fair die'' regardless of the initial 
energy and the launching method. In other words, it presents the same probabilities 
of falling for all faces (provided that collisions are not induced in a specific 
region, altering the multiplicity factors). An example of this is RPG dice, which 
are Platonic solids and exhibit this symmetry, including, of course, the case of the 
traditional six-sided die.

Last but not least, it is worth to mention that this investigation was 
completed as part of the authors' participation in the International Young 
Physicists' Tournament (IYPT), a competition that seeks to encourage 
high school students to solve open physics problems which consist of small 
paragraphs defining a specific situation or phenomenon, and then establish 
some task that will not have a final or closed answer but 
will lead students to find creative and deep explanations for that situation. 
Ordinary high school physics will certainly not be enough to accomplish those 
tasks and, therefore, those students will learn much more than what is usually 
taught in regular curricula. These are, therefore, typical characteristics of an 
active learning method.

\begin{acknowledgements}
We would like to thank  Prof. Silvio R. A. Salinas from the University of São Paulo 
for his supportive opinions and discussion on the statistical mechanics of 
the three-sided dice.
\end{acknowledgements}





\end{document}